\documentclass[12pt]{article}
\usepackage{amsmath}
\usepackage{graphicx,psfrag,epsf}
\usepackage{enumerate}
\usepackage{natbib}
\usepackage{algorithm}%Pseudo code
\usepackage{algorithmic}%Pseudo code
\usepackage{float} %floating images
\usepackage{url} % not crucial - just used below for the URL 
\usepackage{array} %for centerred tables

\newcommand{\blind}{0}

\begin{document}

\def\spacingset#1{\renewcommand{\baselinestretch}%
{#1}\small\normalsize} \spacingset{1}

%%%%%%%%%%%%%%%%%%%%%%%%%%%%%%%%%%%%%%%%%%%%%%%%%%%%%%%%%%%%%%%%%%%%%%%%%%%%%%

\if0\blind
{
  \title{\bf Optimized Spatial Partitioning via Minimal Swarm Intelligence}
  \author{Casey Kneale, Dominic Poerio, Karl S. Booksh\hspace{.2cm}\\
    Department of Chemistry and Biochemistry\\
    University of Delaware Newark Delaware 19716\\
	}
  \maketitle
} \fi

\if1\blind
{
  \bigskip
  \bigskip
  \bigskip
  \begin{center}
    {\LARGE\bf Optimized Spatial Partitioning via Minimal Swarm Intelligence}
\end{center}
  \medskip
} \fi

\bigskip
\begin{abstract}
Optimized spatial partitioning algorithms are the corner stone of many successful experimental designs and statistical methods. Of these algorithms, the Centroidal Voronoi Tessellation (CVT) is the most widely utilized. CVT based methods require global knowledge of spatial boundaries, do not readily allow for weighted regions, have challenging implementations, and are inefficiently extended to high dimensional spaces. We describe two simple partitioning schemes based on nearest and next nearest neighbor locations which easily incorporate these features at the slight expense of optimal placement. Several novel qualitative techniques which assess these partitioning schemes are also included. The feasibility of autonomous uninformed sensor networks utilizing these algorithms are considered. Some improvements in particle swarm optimizer results on multimodal test functions from partitioned initial positions in two space are also illustrated. Pseudo code  for all of the novel algorithms depicted here-in is available in the supplementary information of this manuscript.
\end{abstract}

\noindent%
{\it Keywords:} Centroidal Voronoi Tessellation, Weighted, Obstacle, Autonomous.
\vfill
\hfill {\tiny technometrics tex template (do not remove)}

\newpage
\spacingset{1.45} % DON'T change the spacing!
\section{Introduction}
\label{sec:intro}
Optimized partitioning methods have become of invaluable for aiding in the discovery of heuristic solutions to NP-hard optimization problems, unsupervised classification methods  \citep{Lloyd:57}, experimental design \citep{Kennard:69}, and various implementations of physical technologies. Of the techniques known for optimally partitioning space, the Centroidal Voronoi Tessellation (CVT) is the most widely used. A CVT can be defined by the placement of N points which are the barycenters of N convex partitions whose borders equipartition the distance between each point \citep{Du:99}. The use of CVTs stems from the fact that they have been proven to satisfy the Gersho conjecture for ideal vector quantizers in at least 1 and 2 space \citep{Du:05}. Its use for partitioning has been found especially effective for the aforementioned applications despite the lack of mathematical rigor for higher dimensional spaces.

The computation of Voronoi tessellations requires global knowledge of the search space. This is especially true if the space consists of complicated geometries and bounds. For high dimensional spaces, CVT computations are typically intensive due to the requirement of convex hull algorithms and indistinguishability of simplices. Furthermore, CVT algorithms do not natively support spatial weighting. It must be stated that many approximations and adaptations to the CVT method have been found which incorporate these ideas. However, the merit of this work presented herein is the abandonment of the CVT paradigm and discovery of minimal partitioning approaches which accommodate the aforementioned challenges.

Two simple and effective algorithms for spatial partitioning based on minimal swarm intelligence are detailed herein. These algorithms can be seen to perform similar tasks as CVTs but have different constraints which render them better suited to alternative physical and in silico applications. The utility of these algorithms for weighted regions, nonconvex bounds/obstacles, and higher spatial dimensions are described and assessed. 

Several techniques which characterize the efficacy of these algorithms are employed. In general these assessments are qualitative and exploratory as very few methods which characterize algorithms of this nature currently exist. This preliminary work is concluded with suggestions to improve performance of the aforementioned algorithms, several realized applications, and some open problems.

\section{Description of the Algorithms}
\label{sec:desc}
The algorithms are aptly initialized by normalizing the mean NN distance between randomly generated agents subject to a desired threshold. The primary mechanism by which the algorithm relocates point agents to maximally distant locations with respect to one another is via repelling nearest neighbors (NN). The thresholds for repulsion are set by the mean $L_{1}$ distances between each point and their respective NN. The magnitude of displacement for two points which violate this threshold are based on the difference between their distance and the swarm mean. Eg: $\vec{displacement} = (\bar{\mathbf{d}} - |\mathbf{i} - \mathbf{j}|) \circ (\hat{\mathbf{ij}}/2.0)$ , where $\bar{\mathbf{d}}$ is the mean $L_{1}$ distances for all point\textsc{\char13}s nearest neighbors, $\mathbf{i}$ and $\mathbf{j}$ are the row/column vectors of the point locations.

Such a displacement can be seen as a penalty for closeness. This method is similar to the manner in which smoothed particle hydrodynamic techniques enforce viscosity; sans kernel derivatives and leapfrog integration. However, the driving force for this algorithm is that the swarm mean is increased at each iteration. Agents are thus pushed away from one another based on the swarm\textsc{\char13}s growing nearest neighbor separation until boundary enforcement takes place.

The constraint of spatial boundaries are enforced by an analogy to the method of image charges (Figure 1). If a point is within a mean distance from a boundary, then a projection of that point is constructed across it. Thus, any point approaching a boundary is held in place by its own displacement from each bound by the same penalties enforced for nearest neighbors.

\begin{figure}[H]
\begin{center}
\includegraphics[width=2in]{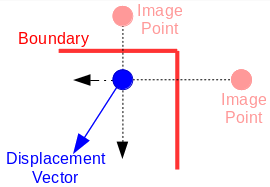}
\end{center}
\caption{An agent (darkly colored circle) near two boundaries (darkly colored rectangles) is repelled by its displacement from image points (lightly colored circles) via vector decomposition. \label{fig:first}}
\end{figure}
Without any other constraints or penalties, the aforementioned algorithm depicts a Repulsive Agent Optimizer (RAO). A secondary repulsion mechanism may also be implemented such that the optimum agent spacing is not hindered by collinear confinement which is enforced by both the boundaries and RAO optimization scheme.

The repulsion scheme which mitigates collinear clustering is based on the nearest and next nearest neighbors. This algorithm creates a line segment from the two nearest neighbors and calculates the nearest perpendicular distance to the respective agent. Distances which are smaller then the mean swarm orthogonal distance are projected away from the nearest neighbors line segment by a magnitude proportional to their difference (Figure 2). 

\begin{figure}[H]
	\begin{center}
		\includegraphics[width=2in]{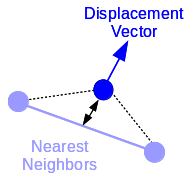}
	\end{center}
	\caption{Orthogonal displacement for any given agent (darkly colored circle) is performed by the projection from the nearest point on the line segment between its nearest and next nearest neighbors(lightly colored circles). \label{fig:second}}
\end{figure}

The RAO and orthogonal nearest neighbors repulsive agent optimizer (ONNRAO) may reach convergence under the same conditions. The convergence criterion is similar to that of CVT, in that a sum of square differences for the agent locations at their previous and current iterations are lower than a predetermined threshold. The displacement criterion isn\textsc{\char13}t always realistic for a human time scale due to chaos/Markov effects of the n-body problem. Thus, a maximum iteration number should always be used as a secondary condition. For most applications such a criterion is implicit.

\section{Comparison of RAO algorithms to CVT}
\label{sec:Comparisons}
Two space provides adequate ground for testing novel space partitioning algorithms. This is because the Gersho conjecture has proven that an optimal vector quantizer which partitions euclidean 2 space should present hexagonal Delauney duals \citep{Du:05}. The RAO and ONNRAO algorithms were tested against CVT in a square bound with 100 agents to assess if the Gersho conjecture appeared to be satisfied under the same conditions (Figure 3).

\begin{figure}[H]
	\begin{center}
		\includegraphics[width=2.75in]{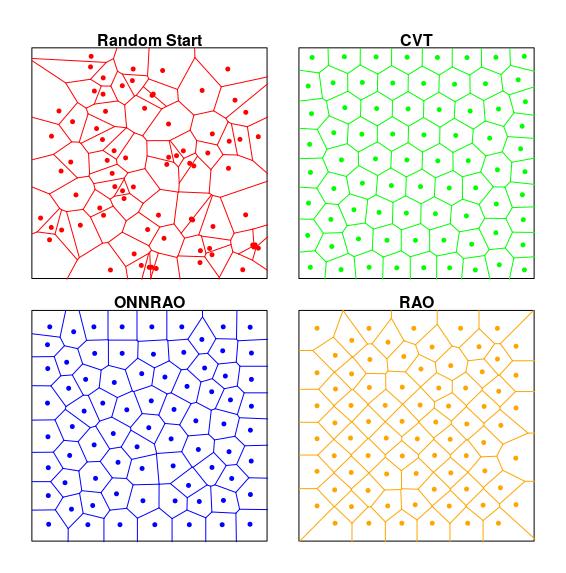}
	\end{center}
	\caption{Agents under Gersho conditions (100 agents) and their corresponding duals optimized from their random starting position (top-left) by the CVT (top-right), ONNRAO (bottom-left), and RAO (bottom-right). \label{fig:third}}
\end{figure}

In the square bound test case, it can be seen that the CVT largely produced hexagon shaped duals. ONNRAO tended to create a variety of spatially inefficient polygons relative to CVT. RAO primarily resulted in quadralateral duals. It appeared that RAO was both the least efficient and the most effected by boundary enforcement of the three. Thus, the Gersho conjecture was not satisfied by either of the RAO algorithms and their partitioning should not be considered optimal. Although the spacings from the RAO algorithms were not optimal, the coefficients of variation for the nearest neighbor and next nearest neighbor distances were always lower then the randomly placed agent trials. 

A statistical test was devised in order to assess if optimal partitioning had taken place under the same conditions. The test created 10,000 randomly located circles or squares of a predescribed area ($0.01^2$ sq units). The number of points which were found in each circle/square were summed and difference of mean tests were performed. The CVT was the most likely to have the same expectation value (1 agent), and had the lowest variance across all trials. RAO typically had the largest variance, and ONNRAO was intermediate. This was not always the case due to the random nature of these algorithms and the parameters used. An auxiliary test outside of Gersho conditions (25 agents) in the same bound was conducted (Figure 4).

\begin{figure}[h]
	\begin{center}
		\includegraphics[width=3in]{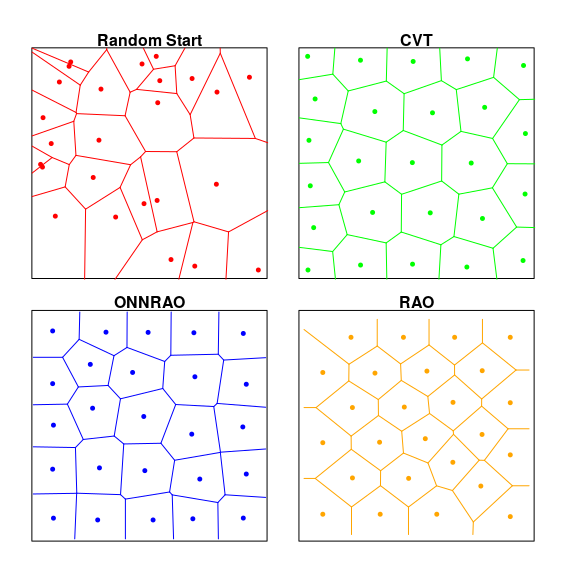}
	\end{center}
	\caption{A 25 agent population test of the three algorithms from the same random starting locations. \label{fig:fourth}}
\end{figure}

It was hypothesized that inside of a square bound an optimal partitioning for a square number of agents would feature diagonals and sides that possess the square root of the number of agents. ONNRAO achieved these conditions in every instance tested. Interestingly, the CVT always featured a square root diagnal, but did not always afford root sides. Only under incredibly rare initial point placements did the RAO achieve either of these conditions in partiality and never in full. The collinear stacking between agents and boundary enforcement at corners likely effects RAO's performance in this regard. The merit of ONNRAO can be seen in that it projects agents orthogonally from collinear nearest neighbors.

A probabilistic study was undertaken to ascertain the likely locations for partitioned agents (CVT, ONNRAO, and RAO) from the same random starting positions. The experiment examined the configurations which resulted from many trials ($>$1000) for 10, 13, and 25 randomly generated points. The space was discretized in to square bins (100x100). A probability for an agent residing in each bin was calculated by a running total obtained from a representative number of algorithm executions and the respective agent populace (Figure 5). Although this test did not provide the infinite perspective of agents placements that differential calculus may afford, it does serve as an empirical representation.

\begin{figure}[h]
	\begin{center}
		\includegraphics[width=5.0in]{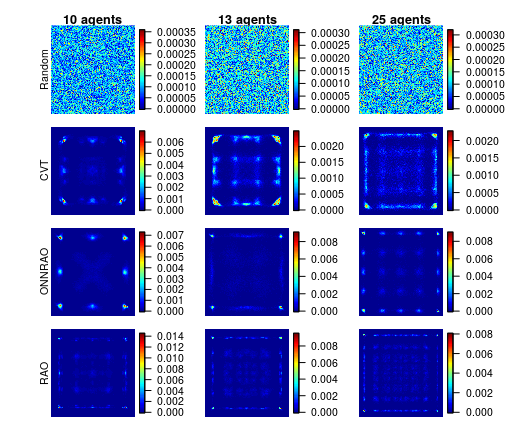}
	\end{center}
	\caption{Heat maps which depict the discretized probability density functions for Random (top), CVT, ONNRAO, and RAO (bottom) point placement for 10 (left), 13, and 25 (right) agents. \label{fig:fifth}}
\end{figure}

From this assessment some general similarities between the algorithms were then observed. For a symmetrically bounded region, a symmetric probability density function (PDF) of optimized partitions ensures that any bias present in the algorithms is uniform. All three of the partitioning algorithms provided symmetric PDFs. The integrated likelihood for the corner regions across all partitioning schemes were of the same magnitude. However, both RAO and ONNRAO strongly favored certain discretized corner locations (4-7x maximum likelihood relative to CVT). Several other qualitative differences of the estimated PDFs should also be mentioned.

The CVT and RAO algorithms tend to make globular clusters of likelihood. However, RAO was observed to result in suboptimal artifacts due to the interference of collinear clustering. Of the three algorithms, ONNRAO produced the most diffuse distributions for tests which contained non-square agent counts. At large agent populations, the ONNRAO PDF most closely resembled those of the CVT. Due to the diffuse positioning in ONNRAO, an examination of the rate for which each algorithm filled the discretized space was pursued.

Another test was conducted for 10, 13, and 25 agent populations in a square bound (30 trials for each condition). The final discretized locations for each agent were stored after each execution. The algorithms were executed until a representative amount of the percent area had been visited by agents. A metric representative of a cumulative distribution function (CDF) was then obtained as a function of algorithm executions for each respective configuration (Figure 6).

\begin{figure}[H]
	\begin{center}
		\includegraphics[width=5.15in]{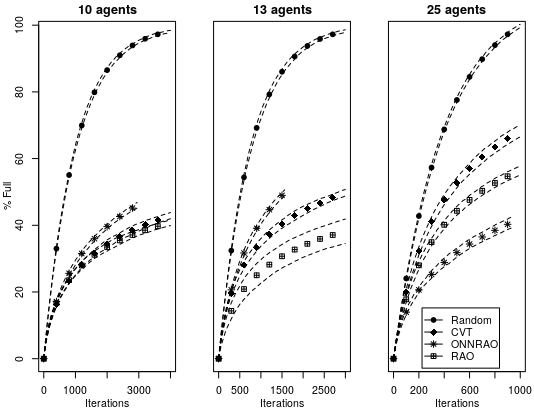}
	\end{center}
	\caption{Plots of discretized area filled verses algorithm executions. Two sigma error margins are depicted by the black dotted lines. \label{fig:sixth}}
\end{figure}

Trends relating to the randomness of the PDF\textsc{\char13}s were obtained from plots of discretized area filled vs algorithm executions. For nonsquare agent populations (10, 13) the ONNRAO had the most freedom to locate agents. Yet when there was a square number of agents, the ONNRAO algorithm became the most restricted of the four. Interestingly, RAO could not achieve uniform spacing for the trial which had a prime number of agents (13) and possessed the most variance. With the exception of random placement, ONNRAO had the least variance in its approximated CDFs. These traits were illustrative of the strengths and weaknesses of the algorithms\textsc{\char13} ability to fill space. 

\section{Weighted Regions}
\label{sec:Wts}
Unlike traditional CVT algorithms, the RAO partitioning schemes are amenable to the implementation of weighted spatial regions. The suggested implementation starts by first performing a test to determine whether a point and/or its interactor are within a given boundary. If either are within the bounds, then the displacement vector is multiplied by the prescribed weight of the region. It must be mentioned that for ONNRAO, the orthogonal projection vector was also weighted in this manner. In principle the implementation of these regions should be effective, however, some nuances were present for RAO.

For proof of concept, several simple tests were performed with heterogeneous circularly weighted regions in 2 space (Figure 7). Both the ONNRAO and RAO were subjected to first a single region (weight = 100) in the middle of a square space, and then to the equal but opposite regions (weight = 200, and 1/200). 

\begin{figure}[h]
	\begin{center}
		\includegraphics[width=2.75in]{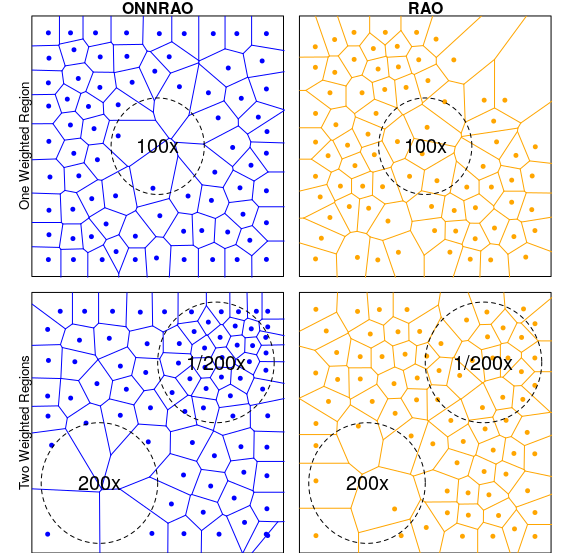}
	\end{center}
	\caption{Comparisons of the ONNRAO and RAO algorithms with circularly weighted regions. The regions are depicted with dashed lines and the weights are inscribed (one weighted region: 100x, two weighted regions: 200x and 1/200x). The same set of randomly generated points were used for each trial. \label{fig:seventh}}
\end{figure}

Large defects were present in the RAO algorithm after the implementation of singularly weighted regions. It was hypothesized that the weighted displacements in RAO often become irreconcilable at early iterations and much of the available area remains empty. Further evidence for the proposed mechanism of failure was observed via the placement of two equal but oppositely weighted regions. If the issue was iterated discrepancies of mean displacement then equal but opposite weights should afford a uniform agent distribution. Observations were made which confirmed this assumption.
 
This issue with RAO could be corrected by applying weights to the regions post convergence of the unweighted space. However, such an implementation adds computational complexity and can be seen as suboptimal relative to ONNRAOs success. Over 100 trials with weighted regions demonstrated that ONNRAO partitioned the space with fewer defects and was more responsive to the assigned weights. Similar results were obtained with heterogeneous obstacles. 

\section{Obstacles}
\label{sec:Obs}
Tests of both RAO algorithms were performed to assess their efficacy for nonconvex boundaries and obstacles. Two simple cases were examined for 45 agents. These cases consisted of a centered square obstacle and that of an offset rectangular obstacle (Figure 8). Both algorithms successfully avoided the obstacles due to the image charge boundary enforcement, but ONNRAO most effectively partitioned the space. Again, RAO was often unable to correct for the loss of available area. 

It should be mentioned that when the agent population was less than 10, the spatial efficiency for both algorithms was hindered. The correction of this issue is an area open to future research. 
 
\begin{figure}[H]
	\begin{center}
		\includegraphics[width=3in]{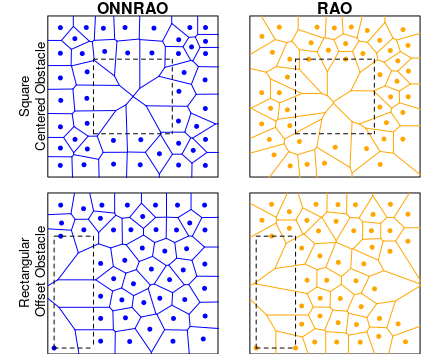}
	\end{center}
	\caption{A plot of rectangular obstacles and their effect on ONNRAO and RAO. A small programming error associated with random initial point placement can be seen in the rectangular offset trials by the agents located at the extreme bottom left locations. \label{fig:eigth}}
\end{figure}

\section{Extensibility to Higher Dimensions}
\label{sec:HiDim}

Similar to CVT, RAO and ONNRAO are both applicable to higher dimensions of euclidean space. RAO and ONNRAO do not require the calculation of N dimensional simplices, volumes, or otherwise. Unfortunately the authors could not compare the results of the RAO algorithms for the higher dimensional tests because an error free implementation of high order CVT could neither be attained nor obtained. A simple test was performed for assessing the efficacy of the RAO algorithms in 2-5 space (Figure 9).

\begin{figure}[H]
	\begin{center}
		\includegraphics[width=4in]{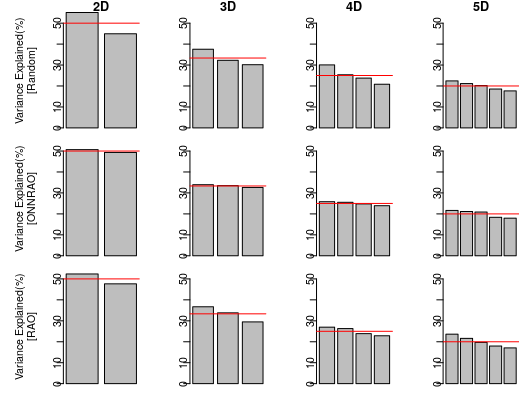}
	\end{center}
	\caption{Scree plots of random(top), ONNRAO(middle), and RAO(bottom) point placement for 2-5 dimensions of space. The expected value for each trial is depicted with the horizontal line. \label{fig:ninth}}
\end{figure}

Principal component analysis (PCA) was applied to randomly, ONNRAO, and RAO generated agent locations. The number of agents for 2, 3, 4, and 5 dimensions were empirically chosen to be 70, 75, 90, and 115, respectively. PCA generates N component vectors which describe the directions of maximum variance for a given set of data. In this case, the variance explained by the principal components can illustrate whether optimized partitioning has in fact occurred.

For an optimized partitioning, the expected value for any principal components\textsc{\char13} explained variance should be 1/N. This is because an optimized partitioning equally weights agent locations by each N dimension of space. Although there is some inherent bias present on random noise redistribution to principal components (1st with the least, and last with the most), the condition should still hold. From this test it was found that ONNRAO was the most robust to higher dimensions. RAO only performed slightly better than randomly generated agent locations for most trials. The primary condition for success of the ONNRAO algorithm appeared to be the agent population, but the number of iterations were also critical.
 
\section{Performance Considerations}
\label{sec:Consid}
The performance of the RAO algorithms is dependent upon the number of maximum iterations, number of agents, and the dimensions of space. The number of iterations required to achieve optimized partitioning for the RAO and ONNRAO are vastly different. For example, with 75 agents confined to a 1x1 unit square, RAO requires $\approx$8500 iterations and ONNRAO only $\approx$600. Thus direct comparisons for performance cannot be made, but general trends related to iteration cost were assessed.

The relative speed of each algorithm to complete 1000 iterations for 2, 3, 4, and 5 dimensional spaces possessing 5, 10, 25, 50, and 75 agents was assessed (Figure 10). It was shown that both algorithms follow exponential time by agent, but only linear time by the dimension of space. RAO iterations were performed faster then ONNRAO ($\approx$4/3x), but again RAO is an order of magnitude less efficient than ONNRAO.

\begin{figure}[H]
	\begin{center}
		\includegraphics[width=4in]{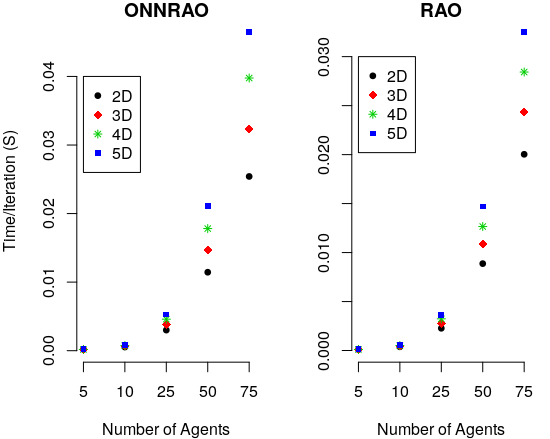}
	\end{center}
	\caption{Plots of time per iteration verses number of agents for ONNRAO (left) and RAO (right). Expense of each iteration for dimensions of space are depicted by different point shapes.\label{fig:tenth}}
\end{figure}

The most effective means to improve performance and memory over-head of these algorithms is by the employment of quadtrees, octrees, k-d trees, and similar space quantization methods. These methods reduce the number of distance assessments between agents and can bring the exponential time dependence on number of particles closer to linear time \citep{Bentley:75}. In physical applications, distance assessments can be made via signal strength and could be acquired more cheaply.

\section{Physical Applications}
\label{sec:PhysApplications}
Perhaps the greatest physical application envisioned for the RAO algorithms is that of uninformed sensor networks. Terrains such as alien planets, outerspace, collapsed infrastructure, or underwater caves make exploration difficult, expensive, or impossible to predetermine the positions of physical boundaries prior to entry. This is especially true if the bounds are dynamic due to the following: erosion, impact events, moving instrumentation or physicochemical properties (pH, chemical signature, temperature, etc). In order to optimally search challenging terrains, sensors must rely on rapid swarm intelligence for efficacy and survival.

The implicit advantage of RAO algorithms for this application is that they only require each sensor to be aware of their individual immediate boundaries, and be able to establish communication with two of their neighbors. With this limited information, autonomous and dynamic navigational decisions which lead to nearly optimal placement can be made in a vector oriented manner. These considerations cannot be directly performed with CVTs. 

Many have attempted to harmonize CVT placements on assumed unperturbed quadrilateral regions with various vector positioning algorithms. A good example of such an effort is that of the combination Lloyd's algorithm and the TangentBug algorithm \citep{Breitenmoser:10} . However, these methods are computationally expensive, often utilize central control, and require more perimeter knowledge than the RAO algorithms. Perhaps most importantly, the code for these algorithms is only a few hundred lines of basic matrix algebra. Microcomputers and even microcontrollers could store and execute the partitioning schemes detailed herein without difficulty. 

\section{Particle Swarm Optimizers}
\label{sec:CompApplications}

Particle swarm optimizers (PSO) were first described by Kennedy, and Eberhart \citep{eberhart:05}. In essence PSO's minimize/maximize an objective function by evaluating it at each position in the function space where a particle has been defined. At every iteration particles change their location based on both the best overall and individual solution until a stopping criterion is met. Much research has focused on the exploration of parameters which affect the degrees of attraction and particle velocities in order to obtain optimal solutions. For simplicity and reproducibility, we have chosen standard PSO implementations and utilized default parameters except where specified.

It has been shown that the optimization of initial particle placements can lead to more efficient and/or accurate PSO outcomes for higher dimensions of space ($>$10) \citep{Rich:04}. Yet, results performed in lower spaces have been left unpublished. Several 2 space tests were devised which served to compare the results of PSO experiments that had initial point locations which were generated by RAO, ONNRAO, CVT, and random placement. 

First a broad survey of many standard optimization test functions were assessed (Supplementary Table 1). These functions were examined with only 10 agents/particles to determine if any of the starting methods increased their performance. It was shown that with a standard PSO2011 algorithm \citep{Mauricio:13} the number of functional evaluations (NFE) was hardly impacted by any of the point placement methods. Over-all the rate of meeting the objective function with-in tolerance (success rate) and efficiencies were commensurate with the exception of a few cases. For the Rosenbrock function, success rates obtained by ONNRAO placement were 7-8\% greater than random placement. Interestingly, CVT and ONNRAO placements allowed for the PSO to minimize the Ackley function 3 orders of magnitudes closer to the objective value (0).

Many of the functions utilized for these tests which demonstrated improvement were multimodal. The presence of local minima provides locations for which PSO can falsely converge. Thus the distance from the known global minima and observed minima were examined for the Rosenbrock, Griewank, and Schwefel functions. In this case, the efficacy of the PSO algorithm \citep{Gilli:11} was assessed by limiting the number of iterations (100) but the swarm consisted of 50 particles. Additionally each PSO was repeated 1000 times for a given starting configuration because it is a stochastic process (Table 1). We report the average absolute error $\epsilon$ for two dimensional space as,
$\epsilon = \frac{|\overline{X} - \mu_{_{X}}| + |\overline{Y} - \mu_{_{Y}}|}{2} $

\begin{table}[htb]
	\begin{center}
		\caption{Average absolute errors from 1000 PSO runs using each point placement method as starting coordinates} \label{tab:tabtwo}
		\begin{tabular}{rllll}%rrrrr}
			Method & Rosenbrock & Griewank & Schwefel \\\hline
			\textbf{RAO}  & $2.28 \cdot 10^{-4}$ & $6.56$ & $0.31$ \\
			\textbf{ONNRAO}  & $6.49 \cdot 10^{-5}$ & $0.81$ & $6.02\cdot 10^{-3}$ \\
			\textbf{CVT}  & $6.06 \cdot 10^{-4}$ & $1.33$ & $3.52$ \\
			\textbf{Random}  & $4.48 \cdot 10^{-4}$ & $21.67$ & $14.51$ \\
		\end{tabular}
	\end{center}
	
\end{table}

In all cases tested, ONNRAO obtained the lowest absolute error. CVT outperformed RAO only in the case of the Griewank function. Interestingly, the random start outperformed CVT in the case of the Rosenbrock function. A more indepth study investigating the limiting conditions for PSO and starting placement is warranted but certainly outside of the scope of this work. For proof of concept, several advantages for PSO algorithms with partitioned starting locations have been illustrated.  

\section{Future Work}
\label{sec:Future}

There are many open questions as to the best means to employ these partitioning schemes, and what other schemes may be considered advantageous. The discovery of reliable and memory efficient methods which can identify looping patterns of agents at optimum partitions are an interesting area of future research. Such methods would allow for an empirical convergence criterion rather then reliance on maximum iteration count. ONNRAO uses two simple repulsion mechanisms to partition space. The investigation of accessory attractive and repulsion mechanisms would also be a fruitful area of study. Especially for problems which require low agent populations and highly obstructed or weighted regions. 

\section{Conclusion}
\label{sec:Conclusion}
Two novel algorithms for optimized spatial partitioning based on nearest neighbor and next nearest neighbor repulsion have been presented. Of these two methods, ONNRAO was the most efficient and applicable to weighted regions and obstacles in N dimensional euclidean space. However, both methods were shown to partition space less effectively than CVT via implications from the Gersho conjecture and elementary measures of central tendency. Despite the loss in optimization performance, the ability to effectively partition weighted N space in a vectorized manner from local boundary assessments and only knowledge of two neighbors is unmatched by similar algorithms. 

The ONNRAO partitioning scheme has other unique attributes which may be considered advantageous for the square bounded case. For nonsquare agent populations, ONNRAO, provides the least spatial constraints for where agents are placed. Some swarm optimizers may benefit from the availability of more area surveyed by repeated executions. Interestingly, for a square number of agents, ONNRAO yielded the most spatial constraints and featured PDFs which always possessed the square root diagonal and sides.

Some promising preliminary results for improvements of particle swarm optimizer in 2 space were also observed. PSO results for several multimodal functions possessed smaller distances to their respective global minima than random starting locations for most partitioning schemes. Similarly, some improvements in success rate and PSO result were observed for the Rosenbrock and Ackley functions respectively.

Future efforts directed toward empirical convergence criteria and real-world applications could furnish an exciting new family of partitioning algorithms which do not require expertise in computational geometry to implement.

\section{Acknowledgements}
\label{sec:acknowledge}
The authors would like to thank Vicki Lyn Wallace for her aid in revising the manuscript and fruitful discussions of minimalist movement schemes.

\newpage 
\begin{center}
{\large\bf SUPPLEMENTARY MATERIAL}
\end{center}

\begin{description}

\item[Title:] Pseudo-code for RAO and ONN-RAO Algorithms

\end{description}

\begin{algorithm}
	\caption{Core Algorithm for ONNRAO and RAO}
	\begin{algorithmic}
		\REQUIRE $Points_{D,P}$, toleranceConverge, toleranceNormalization, MaxIters, expandBy, ONNRAO
		\ENSURE $Points_{D,P}$
		\STATE $meanExpand \leftarrow 1$
		\STATE $Points_{D,P} \leftarrow Initialization(Points{D,P}, toleranceNormalization)$
		\STATE $Diff \leftarrow toleranceConverge + 1$
		\WHILE{$iters < MaxIters$}
		\STATE $meanExpand \leftarrow meanExpand + toleranceConverge * expandBy$
		
		\WHILE{$iters < MaxIters$ \AND $Diff > toleranceConverge$}
		
		\STATE $iters \leftarrow iters + 1$
		\IF{$ONNRAO ==$ TRUE}
		\STATE $newPoints_{D,P} \leftarrow ExpandONN(Points_{D,P})$
		\STATE $newPoints_{D,P} \leftarrow ExpandNN(newPoints_{D,P,meanExpand})$
		\ELSE
		\STATE $newPoints_{D,P} \leftarrow ExpandNN(Points_{D,P},meanExpand)$
		\ENDIF
		
		\FOR{$d \in \{1,\dots,D\}$}
		\STATE $Diff \leftarrow Diff + | \sum{ newPoints(d, ) - Points(d, ) } |$
		\ENDFOR
		\STATE $Diff \leftarrow Diff/D$
		
		\STATE $Points_{D,P} \leftarrow newPoints_{D,P}$
		
		\ENDWHILE
		
		\ENDWHILE
	\end{algorithmic}
\end{algorithm}

\newpage 

\begin{algorithm}
	\caption{Initialization}
	\begin{algorithmic}
		\REQUIRE $Points_{D,P}$ and Threshold
		\ENSURE $Points_{D,P}$
		\WHILE{$Diff > Threshold$}
		\STATE $Old \leftarrow Points_{D,P}$
		\STATE $Points_{D,P} \leftarrow AttractOrRepelByMeanDistance(Points_{D,P})$
		\FOR{$d \in \{1,\dots,D\}$}
		\STATE $Diff \leftarrow Diff + | \sum{ Points(d, ) - oldPoints(d, ) } |$
		\ENDFOR
		\STATE $Diff \leftarrow Diff/D$
		\ENDWHILE
	\end{algorithmic}
\end{algorithm}
	
\begin{algorithm}
	\caption{ExpandONN}
	\begin{algorithmic}
		\REQUIRE $Points_{D,P}$
		\ENSURE $Points_{D,P}$
		\STATE //Acquire next and nearest neighbor information
		\STATE $NN_{D+1,P} \leftarrow NearestNeighbors(Points_{D,P})$
		\STATE $NextNN_{D+1,P} \leftarrow NextNearestNeighbors(Points_{D,P})$
		\STATE //Find distance from each agent to the line between nearest neighbors
		\FOR{$p \in \{1,\dots,P\}$}
		\STATE $PointsOnLine( , p) \leftarrow NearestPointOnLine(NN( , p)), NextNB( , p), Points( , p))$
		\STATE $Distances(p) \leftarrow euclidDistance (PointsOnLine(, p), Points( , p))$
		\ENDFOR

		\STATE $meanDist \leftarrow mean(Distances)$
		\STATE //Calculate average orthogonal L1 distances
		\FOR{$d \in \{1,\dots,D\}$}
		\STATE $idealDists(d) \leftarrow  mean(|PointsOnLine( d,  ) - Points( d,)|)$
		\ENDFOR
		
		\FOR{$p \in \{1,\dots,P\}$}
		\STATE //Move points away from line if they are too close
		\IF{$Distances(p) < meanDist$}
		\STATE $unitVec \leftarrow  UnitVector(Points(, p), PointsOnLine(, p), idealDists)$
				
		\STATE $pointDists \leftarrow |Points( , p) - PointsOnLine(, p)|$
		\STATE $nDiff \leftarrow (idealDists - pointDists)$
		\STATE $Points(, p) \leftarrow Points(, p) + nDiff \circ (unitVec/2.0)$
		\ENDIF
		\STATE $Points(, p) \leftarrow  EnforceBoundaries(Points(, p))$
			
		\ENDFOR
				
	\end{algorithmic}
\end{algorithm}

\begin{algorithm}
	\caption{ExpandNN}
	\begin{algorithmic}
		\REQUIRE $Points_{D,P}, meanExpand$
		\ENSURE $Points_{D,P}$
		\STATE //Acquire nearest neighbor information
		\STATE $NN_{D+1,P} \leftarrow NearestNeighbors(Points_{D,P})$
		\STATE //Find distance from each agent to the line between nearest neighbors
		
		\STATE //Calculate average orthogonal L1 distances
		\FOR{$d \in \{1,\dots,D\}$}
		\STATE $idealDists(d) \leftarrow  mean(NN( d+1,  ) - Points( d,))$
		\ENDFOR
		\STATE $idealDists_{D,1} \leftarrow idealDists_{D,1} * meanExpand$
		
		\FOR{$p \in \{1,\dots,P\}$}
		\FOR{$j \in \{1,\dots,P\}$}
		
		\STATE $PointDists \leftarrow |Points( , i) - Points( , j)|$
		
		\IF{$all(PointDists_{D,1} < idealDists_{D,1})$}
		\STATE $unitVec \leftarrow  UnitVector(Points(, p), Points(, j), idealDists)$
		
		\STATE $pointDists \leftarrow |Points( , p) - PointsOnLine(, p)|$
		\STATE $nDiff \leftarrow (idealDists - pointDists)$
		\STATE $Points(, p) \leftarrow Points(, p) + nDiff \circ (unitVec/2.0)$
		\STATE $Points(, p) \leftarrow EnforceBoundaries(Points(, p))$
		
		\STATE $Points(, j) \leftarrow Points(, j) - nDiff \circ (unitVec/2.0)$
		\STATE $Points(, j) \leftarrow EnforceBoundaries(Points(, j))$
		\ENDIF
		\STATE $Points(, p) \leftarrow  EnforceBoundaries(Points(, p))$
		
		\ENDFOR
		\ENDFOR
	\end{algorithmic}
\end{algorithm}

\begin{table}
	\begin{center}
				\caption{PSO2011 Performance metrics for solving 5 standard mathematical functions$($Ackley, Griewank,Parabola, Rastragin, Rosenbrock$)$ with various initial starting positions. The results tabulated result from 30 initial placements based on the same random locations and 30 PSO executions.} \label{tab:tabone}
		\begin{tabular}{rllll}%rrrrr}
			& CVT & ONNRAO & RAO & Random  \\\hline
			\textbf{Ackley}  & (Objective = 0.0) & & & \\
			Results  & 6.726 $\pm$ 2.268 $(10^{-5})$& 6.683 $\pm$ 2.387$(10^{-5})$ & 0.0029 $\pm$ (0.086) & 0.0029 $\pm$ (0.086) \\
			Relative NFE & 0.940 & 0.955 & 0.999 &  1 \\
			Success Rate & 1 $\pm$ 0 & 1 $\pm$ 0  & 0.998  $\pm$ 0.006  & 0.998  $\pm$ 0.006  \\
			Efficiency & 0.983 $\pm$ 0 & 0.983 $\pm$ 0.001 & 0.983 $\pm$ 0.006 & 0.983 $\pm$ 0.006  \\
			%NextFn
			\textbf{Griewank}  & (Objective = 0.0) & & & \\
			Results  &  0.0243 $\pm$ 0.0321 & 0.0234 $\pm$ 0.0313 & 0.0247 $\pm$ 0.0338 & 0.0240 $\pm$ 0.0310 \\
			Relative NFE & 0.934 & 0.834& 1 & 0.894 \\
			Success Rate & 0.194 $\pm$ 0.0690 & 0.1778 $\pm$ 0.0633  & 0.2144 $\pm$ 0.0781 & 0.2022 $\pm$ 0.0753  \\
			Efficiency & 0.1810 $\pm$ 0.06390 & 0.1680 $\pm$ 0.0604 & 0.1976 $\pm$ 0.0754 & 0.1905 $\pm$ 0.0714\\
			%NextFn
			\textbf{Parabola}  & (Objective = 0.0) & & & \\
			Results$(10^{-9})$  & 5.098 $\pm$ 2.851 & 4.967 $\pm$ 2.944 & 5.092 $\pm$ 2.837 & 4.918 $\pm$ 2.849\\
			Relative NFE & 0.993 & 1.00 & 0.991 & 0.992\\
			Success Rate & 1 $\pm$ 0 & 1 $\pm$ 0 & 1 $\pm$ 0 & 1 $\pm$ 0\\
			Efficiency & 0.9847 $\pm$ 0.00027 & 0.9846 $\pm$ 0.00027 & 0.9848 $\pm$ 0.00024 & 0.9847 $\pm$ 0.00026\\
			%NextFn
			\textbf{Rastrigin}  & (Objective = 0.0) & & & \\
			Results  & 0.5661 $\pm$ 0.690 & 0.7024 $\pm$ 0.9575 & 0.6091 $\pm$ 0.7298 & 0.6341 $\pm$ 0.883971  \\
			Relative NFE & 1.0 & 0.895 & 0.958 & 0.977\\
			Success Rate & 0.5089 $\pm$  0.06605 & 0.4856 $\pm$ 0.1391 & 0.4878 $\pm$ 0.11429 & 0.5089 $\pm$ 0.1449\\
			Efficiency & 0.4102 $\pm$ 0.05750 & 0.4008 $\pm$ 0.1203 & 0.3948 $\pm$ 0.09379 & 0.4132 $\pm$ 0.1255\\
			%NextFn
			\textbf{Rosenbrock}  & (Objective = 0.0) & & & \\
			Results  & 0.4296 $\pm$ 0.8029 & 0.4486 $\pm$ 0.8600 & 0.5242 $\pm$ 0.9671 & 0.7309 $\pm$ 1.220 \\
			Relative NFE & 0.893 & 1 & 0.893 & 0.888 \\
			Success Rate & 0.3322 $\pm$ 0.0920 & 0.3744 $\pm$ 0.0989 & 0.3122 $\pm$ 0.1146 & 0.2911 $\pm$ 0.1103\\
			Efficiency & 0.3259 $\pm$ 0.0915 & 0.3646 $\pm$ 0.0971 & 0.3063 $\pm$ 0.1128 & 0.2854 $\pm$ 0.1083\\
		\end{tabular}
	\end{center}

\end{table}

{}

\end{document}